\documentstyle[11pt,aaspp4]{article}

\newdimen\sa  \newdimen\sb
\def\pdeg{\ifmmode $\setbox0=\hbox{$^{\circ}$}\rlap{\hskip.11\wd0 .}$^{\circ}
          \else \setbox0=\hbox{$^{\circ}$}\rlap{\hskip.11\wd0 .}$^{\circ}$\fi}
\def\gtorder{\mathrel{\raise.3ex\hbox{$>$}\mkern-14mu
             \lower0.6ex\hbox{$\sim$}}}
\def\ltorder{\mathrel{\raise.3ex\hbox{$<$}\mkern-14mu
             \lower0.6ex\hbox{$\sim$}}}

\newcommand{\kms}{\mbox{ km~s$^{-1}$}}

\def\PsfigVersion{1.10}
\def\setDriver{\DvipsDriver} 
\ifx\undefined\psfig\else \fi
\let\LaTeXAtSign=\@
\let\@=\relax
\edef\psfigRestoreAt{\catcode`\@=\number\catcode`@\relax}
\catcode`\@=11\relax
\newwrite\@unused
\def\ps@typeout#1{{\let\protect\string\immediate\write\@unused{#1}}}

\def\DvipsDriver{
	\ps@typeout{psfig/tex \PsfigVersion -dvips}
\def\PsfigSpecials{\DvipsSpecials} 	\def\ps@dir{/}
\def\ps@predir{} }
\def\OzTeXDriver{
	\ps@typeout{psfig/tex \PsfigVersion -oztex}
	\def\PsfigSpecials{\OzTeXSpecials}
	\def\ps@dir{:}
	\def\ps@predir{:}
	\catcode`\^^J=5
}

\def\figurepath{./:}

\def\DoPaths#1{\expandafter\EachPath#1\stoplist}
\def\leer{}
\def\EachPath#1:#2\stoplist{
  \ExistsFile{#1}{\SearchedFile}
  \ifx#2\leer
  \else
    \expandafter\EachPath#2\stoplist
  \fi}
\def\ps@dir{/}
\def\ExistsFile#1#2{%
   \openin1=\ps@predir#1\ps@dir#2
   \ifeof1
       \closein1
   \else
       \closein1
        \ifx\ps@founddir\leer
           \edef\ps@founddir{#1}
        \fi
   \fi}
\def\get@dir#1{%
  \def\ps@founddir{}
  \def\SearchedFile{#1}
  \DoPaths\figurepath
}

\def\@nnil{\@nil}
\def\@empty{}
\def\@psdonoop#1\@@#2#3{}
\def\@psdo#1:=#2\do#3{\edef\@psdotmp{#2}\ifx\@psdotmp\@empty \else
    \expandafter\@psdoloop#2,\@nil,\@nil\@@#1{#3}\fi}
\def\@psdoloop#1,#2,#3\@@#4#5{\def#4{#1}\ifx #4\@nnil \else
       #5\def#4{#2}\ifx #4\@nnil \else#5\@ipsdoloop #3\@@#4{#5}\fi\fi}
\def\@ipsdoloop#1,#2\@@#3#4{\def#3{#1}\ifx #3\@nnil 
       \let\@nextwhile=\@psdonoop \else
      #4\relax\let\@nextwhile=\@ipsdoloop\fi\@nextwhile#2\@@#3{#4}}
\def\@tpsdo#1:=#2\do#3{\xdef\@psdotmp{#2}\ifx\@psdotmp\@empty \else
    \@tpsdoloop#2\@nil\@nil\@@#1{#3}\fi}
\def\@tpsdoloop#1#2\@@#3#4{\def#3{#1}\ifx #3\@nnil 
       \let\@nextwhile=\@psdonoop \else
      #4\relax\let\@nextwhile=\@tpsdoloop\fi\@nextwhile#2\@@#3{#4}}
\ifx\undefined\fbox
\newdimen\fboxrule
\newdimen\fboxsep
\newdimen\ps@tempdima
\newbox\ps@tempboxa
\long\def\fbox#1{\leavevmode\setbox\ps@tempboxa\hbox{#1}\ps@tempdima\fboxrule
    \advance\ps@tempdima \fboxsep \advance\ps@tempdima \dp\ps@tempboxa
   \hbox{\lower \ps@tempdima\hbox
  {\vbox{\hrule height \fboxrule
          \hbox{\vrule width \fboxrule \hskip\fboxsep
          \vbox{\vskip\fboxsep \box\ps@tempboxa\vskip\fboxsep}\hskip 
                 \fboxsep\vrule width \fboxrule}
                 \hrule height \fboxrule}}}}
\fi
\newread\ps@stream
\newif\ifnot@eof       
\newif\if@noisy        
\newif\if@atend        
\newif\if@psfile       
{\catcode`\%=12\global\gdef\epsf@start{
\def\epsf@PS{PS}
\def\epsf@getbb#1{%
\openin\ps@stream=\ps@predir#1
\ifeof\ps@stream\ps@typeout{Error, File #1 not found}\else
   {\not@eoftrue \chardef\other=12
    \def\do##1{\catcode`##1=\other}\dospecials \catcode`\ =10
    \loop
       \if@psfile
	  \read\ps@stream to \epsf@fileline
       \else{
	  \obeyspaces
          \read\ps@stream to \epsf@tmp\global\let\epsf@fileline\epsf@tmp}
       \fi
       \ifeof\ps@stream\not@eoffalse\else
       \if@psfile\else
       \expandafter\epsf@test\epsf@fileline:. \\%
       \fi
          \expandafter\epsf@aux\epsf@fileline:. \\%
       \fi
   \ifnot@eof\repeat
   }\closein\ps@stream\fi}%
\long\def\epsf@test#1#2#3:#4\\{\def\epsf@testit{#1#2}
			\ifx\epsf@testit\epsf@start\else
\ps@typeout{Warning! File does not start with `\epsf@start'.  It may not be a PostScript file.}
			\fi
			\@psfiletrue} 
{\catcode`\%=12\global\let\epsf@percent=
\long\def\epsf@aux#1#2:#3\\{\ifx#1\epsf@percent
   \def\epsf@testit{#2}\ifx\epsf@testit\epsf@bblit
	\@atendfalse
        \epsf@atend #3 . \\%
	\if@atend	
	   \if@verbose{
		\ps@typeout{psfig: found `(atend)'; continuing search}
	   }\fi
        \else
        \epsf@grab #3 . . . \\%
        \not@eoffalse
        \global\no@bbfalse
        \fi
   \fi\fi}%
\def\epsf@grab #1 #2 #3 #4 #5\\{%
   \global\def\epsf@llx{#1}\ifx\epsf@llx\empty
      \epsf@grab #2 #3 #4 #5 .\\\else
   \global\def\epsf@lly{#2}%
   \global\def\epsf@urx{#3}\global\def\epsf@ury{#4}\fi}%
\def\epsf@atendlit{(atend)} 
\def\epsf@atend #1 #2 #3\\{%
   \def\epsf@tmp{#1}\ifx\epsf@tmp\empty
      \epsf@atend #2 #3 .\\\else
   \ifx\epsf@tmp\epsf@atendlit\@atendtrue\fi\fi}

\chardef\psletter = 11 
\chardef\other = 12

\newif \ifdebug 
\newif\ifc@mpute 
\c@mputetrue 

\let\then = \relax
\def\r@dian{pt }
\let\r@dians = \r@dian
\let\dimensionless@nit = \r@dian
\let\dimensionless@nits = \dimensionless@nit
\def\internal@nit{sp }
\let\internal@nits = \internal@nit
\newif\ifstillc@nverging
\def \Mess@ge #1{\ifdebug \then \message {#1} \fi}

{ 
	\catcode `\@ = \psletter
	\gdef \nodimen {\expandafter \n@dimen \the \dimen}
	\gdef \term #1 #2 #3%
	       {\edef \t@ {\the #1}
		\edef \t@@ {\expandafter \n@dimen \the #2\r@dian}%
		\t@rm {\t@} {\t@@} {#3}%
	       }
	\gdef \t@rm #1 #2 #3%
	       {{%
		\count 0 = 0
		\dimen 0 = 1 \dimensionless@nit
		\dimen 2 = #2\relax
		\Mess@ge {Calculating term #1 of \nodimen 2}%
		\loop
		\ifnum	\count 0 < #1
		\then	\advance \count 0 by 1
			\Mess@ge {Iteration \the \count 0 \space}%
			\Multiply \dimen 0 by {\dimen 2}%
			\Mess@ge {After multiplication, term = \nodimen 0}%
			\Divide \dimen 0 by {\count 0}%
			\Mess@ge {After division, term = \nodimen 0}%
		\repeat
		\Mess@ge {Final value for term #1 of 
				\nodimen 2 \space is \nodimen 0}%
		\xdef \Term {#3 = \nodimen 0 \r@dians}%
		\aftergroup \Term
	       }}
	\catcode `\p = \other
	\catcode `\t = \other
	\gdef \n@dimen #1pt{#1} 
}

\def \Divide #1by #2{\divide #1 by #2} 

\def \Multiply #1by #2
       {{
	\count 0 = #1\relax
	\count 2 = #2\relax
	\count 4 = 65536
	\Mess@ge {Before scaling, count 0 = \the \count 0 \space and
			count 2 = \the \count 2}%
	\ifnum	\count 0 > 32767 
	\then	\divide \count 0 by 4
		\divide \count 4 by 4
	\else	\ifnum	\count 0 < -32767
		\then	\divide \count 0 by 4
			\divide \count 4 by 4
		\else
		\fi
	\fi
	\ifnum	\count 2 > 32767 
	\then	\divide \count 2 by 4
		\divide \count 4 by 4
	\else	\ifnum	\count 2 < -32767
		\then	\divide \count 2 by 4
			\divide \count 4 by 4
		\else
		\fi
	\fi
	\multiply \count 0 by \count 2
	\divide \count 0 by \count 4
	\xdef \product {#1 = \the \count 0 \internal@nits}%
	\aftergroup \product
       }}

\def\r@duce{\ifdim\dimen0 > 90\r@dian \then   
		\multiply\dimen0 by -1
		\advance\dimen0 by 180\r@dian
		\r@duce
	    \else \ifdim\dimen0 < -90\r@dian \then  
		\advance\dimen0 by 360\r@dian
		\r@duce
		\fi
	    \fi}

\def\Sine#1%
       {{%
	\dimen 0 = #1 \r@dian
	\r@duce
	\ifdim\dimen0 = -90\r@dian \then
	   \dimen4 = -1\r@dian
	   \c@mputefalse
	\fi
	\ifdim\dimen0 = 90\r@dian \then
	   \dimen4 = 1\r@dian
	   \c@mputefalse
	\fi
	\ifdim\dimen0 = 0\r@dian \then
	   \dimen4 = 0\r@dian
	   \c@mputefalse
	\fi
	\ifc@mpute \then
		\divide\dimen0 by 180
		\dimen0=3.141592654\dimen0
		\dimen 2 = 3.1415926535897963\r@dian 
		\divide\dimen 2 by 2 
		\Mess@ge {Sin: calculating Sin of \nodimen 0}%
		\count 0 = 1 
		\dimen 2 = 1 \r@dian 
		\dimen 4 = 0 \r@dian 
		\loop
			\ifnum	\dimen 2 = 0 
			\then	\stillc@nvergingfalse 
			\else	\stillc@nvergingtrue
			\fi
			\ifstillc@nverging 
			\then	\term {\count 0} {\dimen 0} {\dimen 2}%
				\advance \count 0 by 2
				\count 2 = \count 0
				\divide \count 2 by 2
				\ifodd	\count 2 
				\then	\advance \dimen 4 by \dimen 2
				\else	\advance \dimen 4 by -\dimen 2
				\fi
		\repeat
	\fi		
			\xdef \sine {\nodimen 4}%
       }}

\def\Cosine#1{\ifx\sine\UnDefined\edef\Savesine{\relax}\else
		             \edef\Savesine{\sine}\fi
	{\dimen0=#1\r@dian\advance\dimen0 by 90\r@dian
	 \Sine{\nodimen 0}
	 \xdef\cosine{\sine}
	 \xdef\sine{\Savesine}}}	      

\def\psdraft{
	\def\@psdraft{0}
}
\def\psfull{
	\def\@psdraft{100}
}

\psfull

\newif\if@scalefirst
\def\psscalefirst{\@scalefirsttrue}
\def\psrotatefirst{\@scalefirstfalse}
\psrotatefirst

\newif\if@draftbox
\def\psnodraftbox{
	\@draftboxfalse
}
\def\psdraftbox{
	\@draftboxtrue
}
\@draftboxtrue

\newif\if@prologfile
\newif\if@postlogfile
\def\pssilent{
	\@noisyfalse
}
\def\psnoisy{
	\@noisytrue
}
\psnoisy
\newif\if@bbllx
\newif\if@bblly
\newif\if@bburx
\newif\if@bbury
\newif\if@height
\newif\if@width
\newif\if@rheight
\newif\if@rwidth
\newif\if@angle
\newif\if@clip
\newif\if@verbose
\def\@p@@sclip#1{\@cliptrue}
\newif\if@decmpr
\def\@p@@sfigure#1{\def\@p@sfile{null}\def\@p@sbbfile{null}\@decmprfalse
   \openin1=\ps@predir#1
   \ifeof1
	\closein1
	\get@dir{#1}
	\ifx\ps@founddir\leer
		\openin1=\ps@predir#1.bb
		\ifeof1
			\closein1
			\get@dir{#1.bb}
			\ifx\ps@founddir\leer
				\ps@typeout{Can't find #1 in \figurepath}
			\else
				\@decmprtrue
				\def\@p@sfile{\ps@founddir\ps@dir#1}
				\def\@p@sbbfile{\ps@founddir\ps@dir#1.bb}
			\fi
		\else
			\closein1
			\@decmprtrue
			\def\@p@sfile{#1}
			\def\@p@sbbfile{#1.bb}
		\fi
	\else
		\def\@p@sfile{\ps@founddir\ps@dir#1}
		\def\@p@sbbfile{\ps@founddir\ps@dir#1}
	\fi
   \else
	\closein1
	\def\@p@sfile{#1}
	\def\@p@sbbfile{#1}
   \fi
}
\def\@p@@sfile#1{\@p@@sfigure{#1}}
\def\@p@@sbbllx#1{
		\@bbllxtrue
		\dimen100=#1
		\edef\@p@sbbllx{\number\dimen100}
}
\def\@p@@sbblly#1{
		\@bbllytrue
		\dimen100=#1
		\edef\@p@sbblly{\number\dimen100}
}
\def\@p@@sbburx#1{
		\@bburxtrue
		\dimen100=#1
		\edef\@p@sbburx{\number\dimen100}
}
\def\@p@@sbbury#1{
		\@bburytrue
		\dimen100=#1
		\edef\@p@sbbury{\number\dimen100}
}
\def\@p@@sheight#1{
		\@heighttrue
		\dimen100=#1
   		\edef\@p@sheight{\number\dimen100}
}
\def\@p@@swidth#1{
		\@widthtrue
		\dimen100=#1
		\edef\@p@swidth{\number\dimen100}
}
\def\@p@@srheight#1{
		\@rheighttrue
		\dimen100=#1
		\edef\@p@srheight{\number\dimen100}
}
\def\@p@@srwidth#1{
		\@rwidthtrue
		\dimen100=#1
		\edef\@p@srwidth{\number\dimen100}
}
\def\@p@@sangle#1{
		\@angletrue
		\edef\@p@sangle{#1} 
}
\def\@p@@ssilent#1{ 
		\@verbosefalse
}
\def\@p@@sprolog#1{\@prologfiletrue\def\@prologfileval{#1}}
\def\@p@@spostlog#1{\@postlogfiletrue\def\@postlogfileval{#1}}
\def\@cs@name#1{\csname #1\endcsname}
\def\@setparms#1=#2,{\@cs@name{@p@@s#1}{#2}}
\def\ps@init@parms{
		\@bbllxfalse \@bbllyfalse
		\@bburxfalse \@bburyfalse
		\@heightfalse \@widthfalse
		\@rheightfalse \@rwidthfalse
		\def\@p@sbbllx{}\def\@p@sbblly{}
		\def\@p@sbburx{}\def\@p@sbbury{}
		\def\@p@sheight{}\def\@p@swidth{}
		\def\@p@srheight{}\def\@p@srwidth{}
		\def\@p@sangle{0}
		\def\@p@sfile{} \def\@p@sbbfile{}
		\def\@p@scost{10}
		\def\@sc{}
		\@prologfilefalse
		\@postlogfilefalse
		\@clipfalse
		\if@noisy
			\@verbosetrue
		\else
			\@verbosefalse
		\fi
}
\def\parse@ps@parms#1{
	 	\@psdo\@psfiga:=#1\do
		   {\expandafter\@setparms\@psfiga,}}
\newif\ifno@bb
\def\bb@missing{
	\if@verbose{
		\ps@typeout{psfig: searching \@p@sbbfile \space  for bounding box}
	}\fi
	\no@bbtrue
	\epsf@getbb{\@p@sbbfile}
        \ifno@bb \else \bb@cull\epsf@llx\epsf@lly\epsf@urx\epsf@ury\fi
}	
\def\bb@cull#1#2#3#4{
	\dimen100=#1 bp\edef\@p@sbbllx{\number\dimen100}
	\dimen100=#2 bp\edef\@p@sbblly{\number\dimen100}
	\dimen100=#3 bp\edef\@p@sbburx{\number\dimen100}
	\dimen100=#4 bp\edef\@p@sbbury{\number\dimen100}
	\no@bbfalse
}
\newdimen\p@intvaluex
\newdimen\p@intvaluey
\def\rotate@#1#2{{\dimen0=#1 sp\dimen1=#2 sp
		  \global\p@intvaluex=\cosine\dimen0
		  \dimen3=\sine\dimen1
		  \global\advance\p@intvaluex by -\dimen3
		  \global\p@intvaluey=\sine\dimen0
		  \dimen3=\cosine\dimen1
		  \global\advance\p@intvaluey by \dimen3
		  }}
\def\compute@bb{
		\no@bbfalse
		\if@bbllx \else \no@bbtrue \fi
		\if@bblly \else \no@bbtrue \fi
		\if@bburx \else \no@bbtrue \fi
		\if@bbury \else \no@bbtrue \fi
		\ifno@bb \bb@missing \fi
		\ifno@bb \ps@typeout{FATAL ERROR: no bb supplied or found}
			\no-bb-error
		\fi
		\count203=\@p@sbburx
		\count204=\@p@sbbury
		\advance\count203 by -\@p@sbbllx
		\advance\count204 by -\@p@sbblly
		\edef\ps@bbw{\number\count203}
		\edef\ps@bbh{\number\count204}
		\if@angle 
			\Sine{\@p@sangle}\Cosine{\@p@sangle}
	        	{\dimen100=\maxdimen\xdef\r@p@sbbllx{\number\dimen100}
					    \xdef\r@p@sbblly{\number\dimen100}
			                    \xdef\r@p@sbburx{-\number\dimen100}
					    \xdef\r@p@sbbury{-\number\dimen100}}
                        \def\minmaxtest{
			   \ifnum\number\p@intvaluex<\r@p@sbbllx
			      \xdef\r@p@sbbllx{\number\p@intvaluex}\fi
			   \ifnum\number\p@intvaluex>\r@p@sbburx
			      \xdef\r@p@sbburx{\number\p@intvaluex}\fi
			   \ifnum\number\p@intvaluey<\r@p@sbblly
			      \xdef\r@p@sbblly{\number\p@intvaluey}\fi
			   \ifnum\number\p@intvaluey>\r@p@sbbury
			      \xdef\r@p@sbbury{\number\p@intvaluey}\fi
			   }
			\rotate@{\@p@sbbllx}{\@p@sbblly}
			\minmaxtest
			\rotate@{\@p@sbbllx}{\@p@sbbury}
			\minmaxtest
			\rotate@{\@p@sbburx}{\@p@sbblly}
			\minmaxtest
			\rotate@{\@p@sbburx}{\@p@sbbury}
			\minmaxtest
			\edef\@p@sbbllx{\r@p@sbbllx}\edef\@p@sbblly{\r@p@sbblly}
			\edef\@p@sbburx{\r@p@sbburx}\edef\@p@sbbury{\r@p@sbbury}
		\fi
		\count203=\@p@sbburx
		\count204=\@p@sbbury
		\advance\count203 by -\@p@sbbllx
		\advance\count204 by -\@p@sbblly
		\edef\@bbw{\number\count203}
		\edef\@bbh{\number\count204}
}
\def\in@hundreds#1#2#3{\count240=#2 \count241=#3
		     \count100=\count240	
		     \divide\count100 by \count241
		     \count101=\count100
		     \multiply\count101 by \count241
		     \advance\count240 by -\count101
		     \multiply\count240 by 10
		     \count101=\count240	
		     \divide\count101 by \count241
		     \count102=\count101
		     \multiply\count102 by \count241
		     \advance\count240 by -\count102
		     \multiply\count240 by 10
		     \count102=\count240	
		     \divide\count102 by \count241
		     \count200=#1\count205=0
		     \count201=\count200
			\multiply\count201 by \count100
		 	\advance\count205 by \count201
		     \count201=\count200
			\divide\count201 by 10
			\multiply\count201 by \count101
			\advance\count205 by \count201
		     \count201=\count200
			\divide\count201 by 100
			\multiply\count201 by \count102
			\advance\count205 by \count201
		     \edef\@result{\number\count205}
}
\def\compute@wfromh{
		\in@hundreds{\@p@sheight}{\@bbw}{\@bbh}
		\edef\@p@swidth{\@result}
}
\def\compute@hfromw{
	        \in@hundreds{\@p@swidth}{\@bbh}{\@bbw}
		\edef\@p@sheight{\@result}
}
\def\compute@handw{
		\if@height 
			\if@width
			\else
				\compute@wfromh
			\fi
		\else 
			\if@width
				\compute@hfromw
			\else
				\edef\@p@sheight{\@bbh}
				\edef\@p@swidth{\@bbw}
			\fi
		\fi
}
\def\compute@resv{
		\if@rheight \else \edef\@p@srheight{\@p@sheight} \fi
		\if@rwidth \else \edef\@p@srwidth{\@p@swidth} \fi
}
\def\compute@sizes{
	\compute@bb
	\if@scalefirst\if@angle
	\if@width
	   \in@hundreds{\@p@swidth}{\@bbw}{\ps@bbw}
	   \edef\@p@swidth{\@result}
	\fi
	\if@height
	   \in@hundreds{\@p@sheight}{\@bbh}{\ps@bbh}
	   \edef\@p@sheight{\@result}
	\fi
	\fi\fi
	\compute@handw
	\compute@resv}
\def\OzTeXSpecials{
	\special{empty.ps /@isp {true} def}
	\special{empty.ps \@p@swidth \space \@p@sheight \space
			\@p@sbbllx \space \@p@sbblly \space
			\@p@sbburx \space \@p@sbbury \space
			startTexFig \space }
	\if@clip{
		\if@verbose{
			\ps@typeout{(clip)}
		}\fi
		\special{empty.ps doclip \space }
	}\fi
	\if@angle{
		\if@verbose{
			\ps@typeout{(rotate)}
		}\fi
		\special {empty.ps \@p@sangle \space rotate \space} 
	}\fi
	\if@prologfile
	    \special{\@prologfileval \space } \fi
	\if@decmpr{
		\if@verbose{
			\ps@typeout{psfig: Compression not available
			in OzTeX version \space }
		}\fi
	}\else{
		\if@verbose{
			\ps@typeout{psfig: including \@p@sfile \space }
		}\fi
		\special{epsf=\ps@predir\@p@sfile \space }
	}\fi
	\if@postlogfile
	    \special{\@postlogfileval \space } \fi
	\special{empty.ps /@isp {false} def}
}
\def\DvipsSpecials{
	\special{ps::[begin] 	\@p@swidth \space \@p@sheight \space
			\@p@sbbllx \space \@p@sbblly \space
			\@p@sbburx \space \@p@sbbury \space
			startTexFig \space }
	\if@clip{
		\if@verbose{
			\ps@typeout{(clip)}
		}\fi
		\special{ps:: doclip \space }
	}\fi
	\if@angle
		\if@verbose{
			\ps@typeout{(clip)}
		}\fi
		\special {ps:: \@p@sangle \space rotate \space} 
	\fi
	\if@prologfile
	    \special{ps: plotfile \@prologfileval \space } \fi
	\if@decmpr{
		\if@verbose{
			\ps@typeout{psfig: including \@p@sfile.Z \space }
		}\fi
		\special{ps: plotfile "`zcat \@p@sfile.Z" \space }
	}\else{
		\if@verbose{
			\ps@typeout{psfig: including \@p@sfile \space }
		}\fi
		\special{ps: plotfile \@p@sfile \space }
	}\fi
	\if@postlogfile
	    \special{ps: plotfile \@postlogfileval \space } \fi
	\special{ps::[end] endTexFig \space }
}
\def\psfig#1{\vbox {
	\ps@init@parms
	\parse@ps@parms{#1}
	\compute@sizes
	\ifnum\@p@scost<\@psdraft{
		\PsfigSpecials 
		\vbox to \@p@srheight sp{
			\hbox to \@p@srwidth sp{
				\hss
			}
		\vss
		}
	}\else{
		\if@draftbox{		
			\hbox{\fbox{\vbox to \@p@srheight sp{
			\vss
			\hbox to \@p@srwidth sp{ \hss 
			 \hss }
			\vss
			}}}
		}\else{
			\vbox to \@p@srheight sp{
			\vss
			\hbox to \@p@srwidth sp{\hss}
			\vss
			}
		}\fi

	}\fi
}}
\psfigRestoreAt
\setDriver
\let\@=\LaTeXAtSign

\def\fluxa{{ergs~s$^{-1}$~cm$^{-2}$}}
\def\fluxb{\hbox{ergs~s}^{-1}\hbox{cm}^{-2}}

\begin{document}

\title{Finding Gravitational Lenses With X-rays}

\vskip 2truecm

\author{J. A. Mu\~noz, C. S. Kochanek \& E. E. Falco } 
\affil{Harvard-Smithsonian Center for Astrophysics, 60 Garden St., Cambridge,
	MA 02138}
\affil{email: jmunoz, ckochanek, efalco@cfa.harvard.edu}

\begin{abstract}

There are $\sim 1$, $0.1$ and $0.01$ gravitationally lensed X-ray
sources per square degree with soft X-ray fluxes exceeding $10^{-15}$, $10^{-14}$
and $10^{-13}$~\fluxa  ~respectively.  These sources will be detected 
serendipitously with the Chandra X-ray Observatory at a rate of 1--3 lenses per 
year of high resolution imaging.  The low detection rate is due to the
small area over which the HRC and ACIS cameras have the $<1\farcs5$ FWHM
resolution necessary to find gravitational lenses produced by galaxies.
Deep images of rich clusters at intermediate redshifts should yield one 
wide separation ($\Delta\theta \gtorder 5\farcs0$) multiply-imaged background
X-ray source for every $\sim 10$, $30$ and $300$ clusters imaged
to the same flux limits.
\end{abstract}

\keywords{cosmology:~theory -- gravitational lensing --
          cosmology:~observations -- quasars:~general -- X-rays:~galaxies}
\section{INTRODUCTION}

Gravitational lenses are an increasingly powerful tool for studies of
cosmology (Falco, Kochanek \& Mu\~noz 1998, Cooray 1999, Helbig 1999),
the Hubble constant (Impey et al. 1998, Barkana et al. 1999, Bernstein
\& Fischer 1999, Fassnacht et al. 1999), galactic structure (Keeton,
Kochanek \& Falco 1998) and galactic evolution (Kochanek et al. 1999).
Their utility is growing at a fast pace because the number of known lenses is
increasing rapidly, having reached $\sim 50$ systems at present (see
http://cfa-www.harvard.edu/castles).  Despite the larger samples, we
have discovered only a small fraction of the total number of lenses
detectable with modern instruments.

Confusion is a fundamental problem for existing gravitational lens
surveys.  Even at high Galactic latitudes, most point sources found
near quasars are stars rather than gravitationally lensed 
images (see Kochanek 1993a). Confusion in radio lens surveys is caused by the range of
source structures -- flat-spectrum radio lens surveys contain far more
compact doubles than two-image lenses (see Helbig et al. 1999), and
steep-spectrum surveys must cope with the enormous variety of extended
radio-emission morphologies (e.g., Griffith et al. 1991).  These problems
vanish for high-resolution X-ray imaging observations, where confusing
Galactic sources are rare (as at radio wavelengths) and source structure is
simple (as for quasars).  However, the image separations in lenses
are small -- of the nearly $50$ known lenses, 90\% are larger than
0\farcs5, the median separation is 1\farcs5, and only 10\% are larger
than 2\farcs5 (Keeton et al. 1998) -- thus, high angular
resolution (of order $1\arcsec$) is required.  The resolution problem
can be avoided by looking for lensed images produced by rich clusters,
where the image separation is much larger (e.g., Luppino et al. 1999).
 
The High Resolution Camera (HRC) and the AXAF CCD Imaging Spectrometer
(ACIS) on the Chandra X-ray Observatory will allow the first direct
searches for gravitational lenses at X-ray wavelengths.  Both HRC and
ACIS combine a relatively wide field of view with high spatial
resolution near the center of the field (50\% enclosed energy radius
$r_{50} \simeq 0\farcs5$). Unfortunately, the resolution worsens with the
distance from the field center $D$, with $r_{50} \simeq 0\farcs5 +
6\farcs0(D/10')^2$ (see Kenter et al. 1997), and only the central
portion of the detector will be useful for recognizing the typical
lensed source.  In \S2 we estimate the probability of lensing X-ray
emitting AGN as a function of flux, including rough estimates of the
observational selection effects for the HRC and ACIS detectors.  In \S3 we
estimate the probability of finding lensed X-ray AGN in fields
centered on massive clusters at intermediate redshift, where the
larger image separations make the worsening resolution at large
off-axis angles relatively unimportant.  We summarize our conclusions
in \S4.

\section{SERENDIPITOUS LENSES}

The method for calculating the expected number of lenses is well
developed; we follow the calculations used for the radio lens
surveys by King \& Browne (1996), Kochanek (1996), Falco, Kochanek \&
Mu\~noz (1998), Cooray (1999) and Helbig et al. (1999).  We assume
that the lens galaxies are described by singular isothermal spheres
(SIS) with parameter normalizations derived from the best fits to the
multiply-imaged radio sources and quasars in Kochanek (1996) and
Falco et al. (1998).  The SIS mass distribution is generally
consistent with the available lens data (see, e.g., Kochanek 1995), as
well as local stellar dynamical measurements (Rix et al. 1997) and X-ray
observations (e.g., Fabbiano 1989) of early-type galaxies.  We ignore
spiral galaxy lenses, as they are a small fraction of all lenses (10--20\%)
and produce $\simeq50\%$ smaller image separations (see Kochanek 1996). 
We describe the early-type lens galaxies by a constant comoving Schechter (1976)
luminosity function
\begin{equation}
   { dn \over dL } = { n_* \over L_* } \left( { L \over L_* } \right)^\alpha \exp(-L/L_*)
\end{equation}
and a Faber-Jackson (1976) relation to convert from luminosity
to velocity dispersion,
\begin{equation}
    { L \over L_*} = \left( { \sigma \over \sigma_* } \right)^\gamma. 
\end{equation}
The parameters $n_*= 0.0061h^3$~Mpc$^{-3}$ , $\alpha = -1.0$ and $\gamma=4$ 
are measured for nearby galaxies, and $\sigma_*=225\kms$ is measured by
fitting the observed separation distribution of lenses.  This 
parameterization represents the ``standard''
model of Kochanek (1996) and Falco et al. (1998).  Recent revisions
to the model suggested by Chiba \& Yoshii (1999) and Cheng \& Krauss (1999) are
generally inconsistent with the observations (see Kochanek et
al. 1999).  The probability that a source lies within the 
multiple-imaging region of a lens, also known as the optical depth, has a
characteristic scale of $\tau_* = 16 \pi^3 (\sigma_*/c)^4 n_*
r_H^3=0.026$ given the parameters for the mass and number of lens
galaxies.  Although the Hubble radius $r_H=c/H_0$ enters the expression
for the optical depth, the quantity $r_H^3 n_*$ is independent of the
value of the Hubble constant.  
In a flat cosmological model, the optical depth is simply
$\tau= (\tau_*/30) (D_{OS}/r_H)^3$, where $D_{OS}$ is the comoving
distance to the source (Turner 1990; see Carroll, Press \& Turner 1992,
Kochanek 1993b for general expressions).  The average
optical depth is closely related to the square of the observed image
separations, with $\tau \sim \langle\Delta\theta\rangle^2 n_*
D_{OS}^3$ for all cosmologies and lens models.  The characteristic
image separation is $\Delta\theta_* = 8\pi (\sigma_*/c)^2=2\farcs92$,
and in a flat cosmological model the average image separation is
simply $\langle \Delta\theta \rangle = \Delta\theta_*/2$.

We use the soft X-ray (0.3--3.5 KeV) luminosity functions derived by
Boyle et al. (1994), particularly their models H (for $\Omega_0=1$) and K (for
$\Omega_0=0$).  For an X-ray luminosity function $d N/ d L dz $, the
total number of unlensed X-ray sources brighter than flux $S$ per
unit solid angle is
\begin{equation}
   { dN \over d\Omega} (>S) = \int_0^\infty dV_s \int_{L_{min}}^\infty d L { d N \over dL d z}(L) 
\end{equation}
where $dV_s$ is the volume element, and $L_{min} = 4\pi D_{lum}^2 S
(1+z)^{\alpha-1}$ is determined from the luminosity distance $D_{lum}$
and the spectral index $\alpha$ defined by $F_\nu \propto
\nu^{-\alpha}$. Boyle et al. (1994) use $H_0=50$~km~s$^{-1}$~Mpc$^{-1}$,
and  assume a fixed spectral index of $\alpha=1$.  In a flat cosmology
the volume element is $dV_s = D_{OS}^2 dD_{OS}$ where $D_{OS}$
is the comoving distance to the source, and the
luminosity distance is $D_{lum} = D_{OS}(1+z_s)$ in all cosmologies.  To find
the number of lensed X-ray sources we must include the redshift-dependent 
optical depth and the magnification bias (see Schneider, Ehlers \& Falco 1992), 
so the number of
lensed X-ray sources in a flat cosmology is
\begin{equation}
   \left({ dN \over d\Omega}\right)_L(>S) = 
       \int_0^\infty dV_s \tau(z_s) \int_{L_{min}}^\infty d L \int_{M_{min}}^\infty 
        { d M \over M } { d P \over d M} { d N \over dL d z} \left( {L \over M } \right) 
       C\left( { \Delta \theta_{min} \over \Delta\theta_*} \right). 
\end{equation}
The number of lensed sources is related to the number of unlensed
sources through the optical depth at a given source redshift
$\tau(z_s)$, the magnification of the lensed sources relative to the
unlensed sources as described by the magnification probability
distribution $dP/dM$, and selection limits on the detectable image
flux ratios and separations.  For the SIS lens the 
probability distribution 
for the magnification is $dP/dM=8/M^3$ and the minimum detectable
flux ratio $f < 1$ sets the lower limit of the magnification integral,
$M_{min} =2 (1+f)/(1-f)$.  We must also eliminate the lenses with
separations below the resolution limit of Chandra. The angular
selection function
\begin{equation}
  C(x=\Delta\theta_{min}/\Delta\theta_*) = 30 \int_0^1 du u^2 (1-u)^2 \exp(-x^2/u^2) 
\end{equation}
gives the fraction of lenses with separations larger than a minimum
value $\Delta\theta_{min}$.  The expressions for the optical depth 
and the volume element change for cosmological models with non-zero 
curvature (see Carroll et al. 1992 and
Kochanek 1993b for general expressions).  We present the results for
the two cosmologies, $\Omega_0=0$ and $\Omega_0=1$, for which the
X-ray LF was derived by Boyle et al. (1994).  The results for the 
$\Omega_0=0$ model should be similar to those for a flat model with 
$\Omega_0=0.5$ and a cosmological constant (see Carroll et al. 1992).

\begin{figure}
\centerline{\psfig{figure=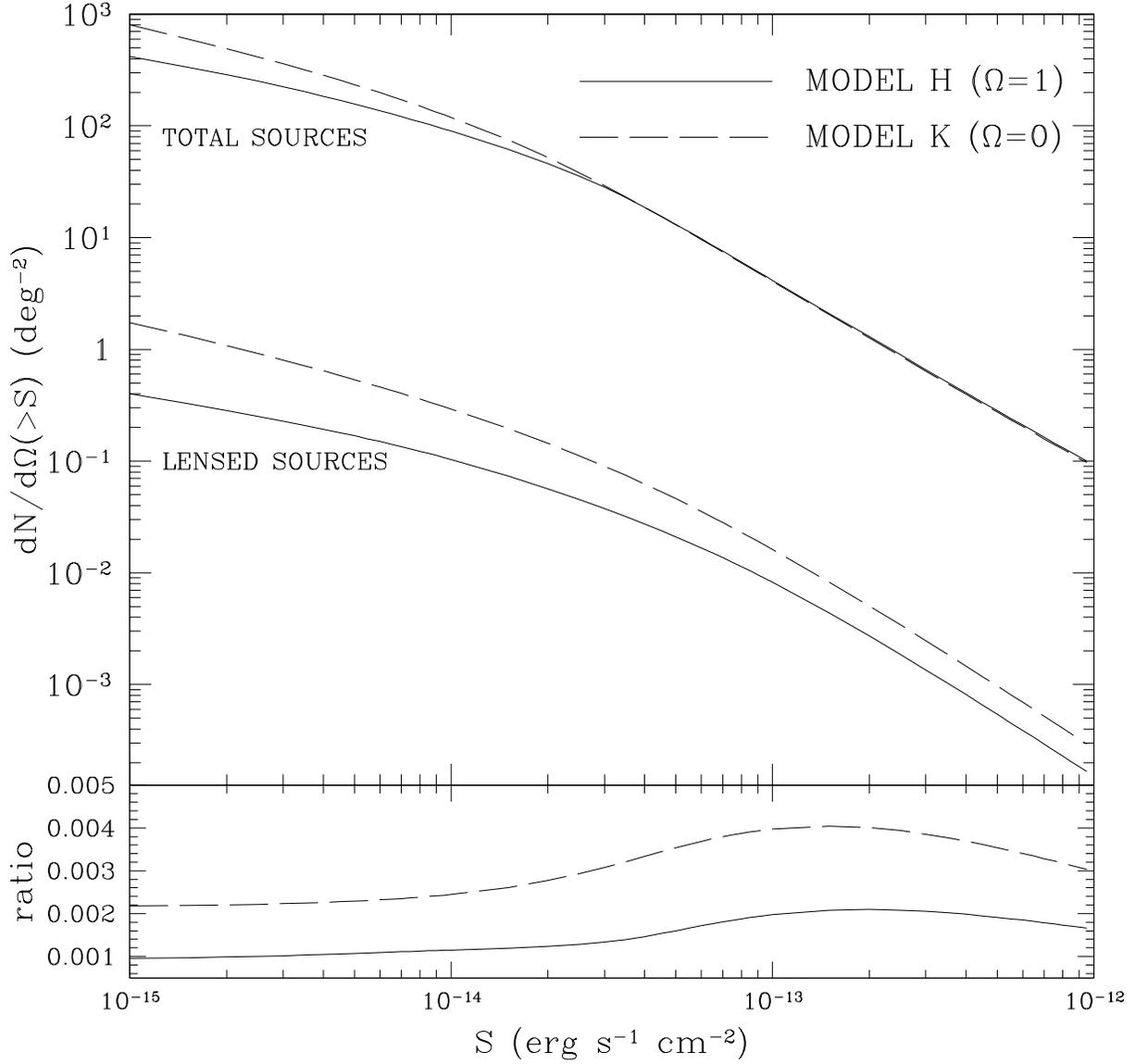,height=6.5in}}
\caption{The number of unlensed (upper curves) and lensed (lower
curves) X-ray sources brighter than flux $S$ per square degree for
either $\Omega_0=1$ and LF model H (solid curves) or $\Omega_0=0$ and
LF model K (dashed curves) from Boyle et al. (1994).  The lower panel
shows the ratios of the curves, which are the fraction of all sources
brighter than flux $S$ that are gravitational lenses.  }
\end{figure}

\begin{figure}
\centerline{\psfig{figure=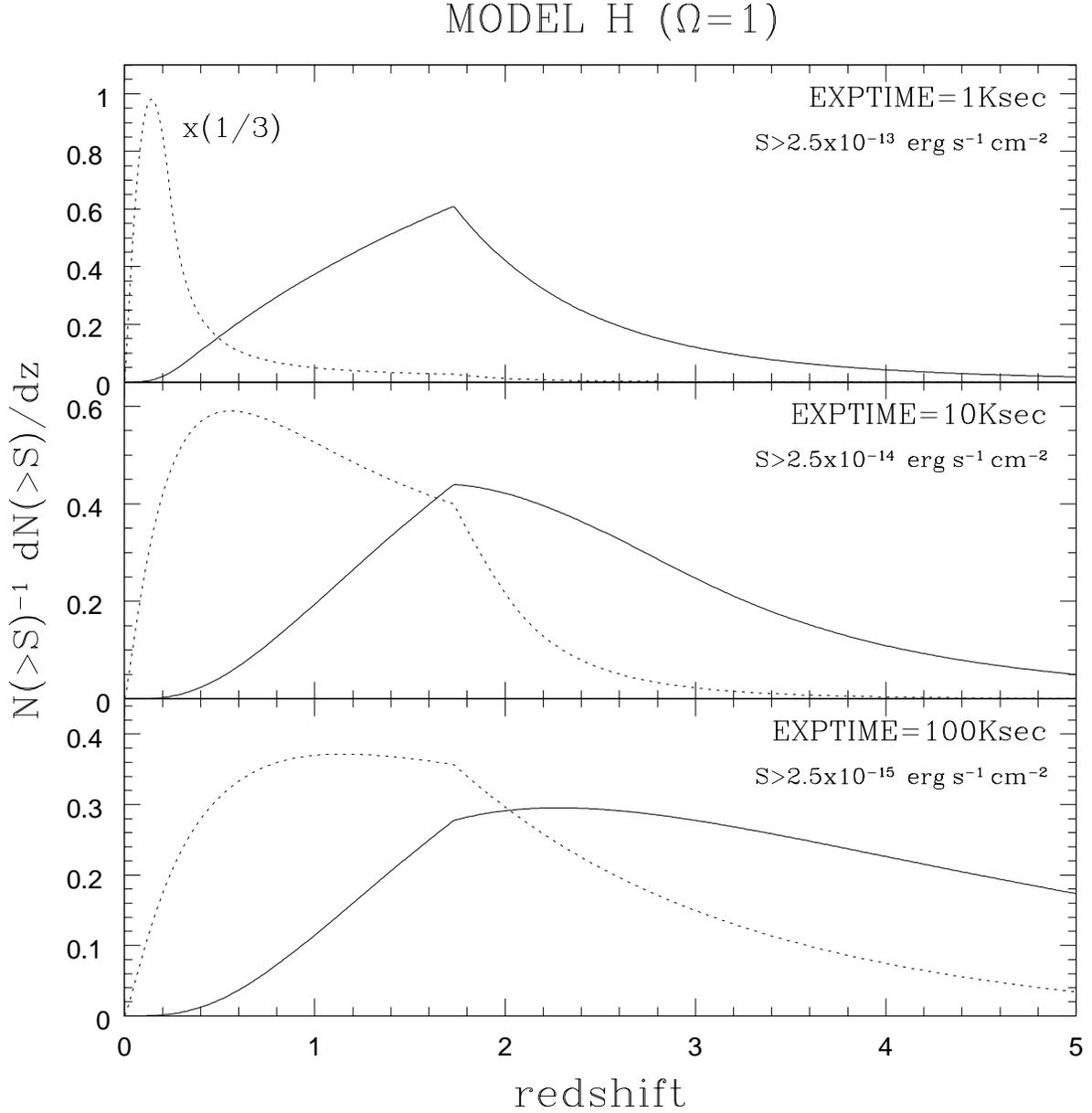,height=6.5in}}
\caption{The normalized redshift distributions of unlensed (dashed)
and lensed (solid) X-ray sources for typical exposure levels of 1, 10,
and 100~ksec for $\Omega_0=1$.  These distributions are independent of
the selection function in flux ratio or separation. }
\end{figure}

\begin{figure}
\centerline{\psfig{figure=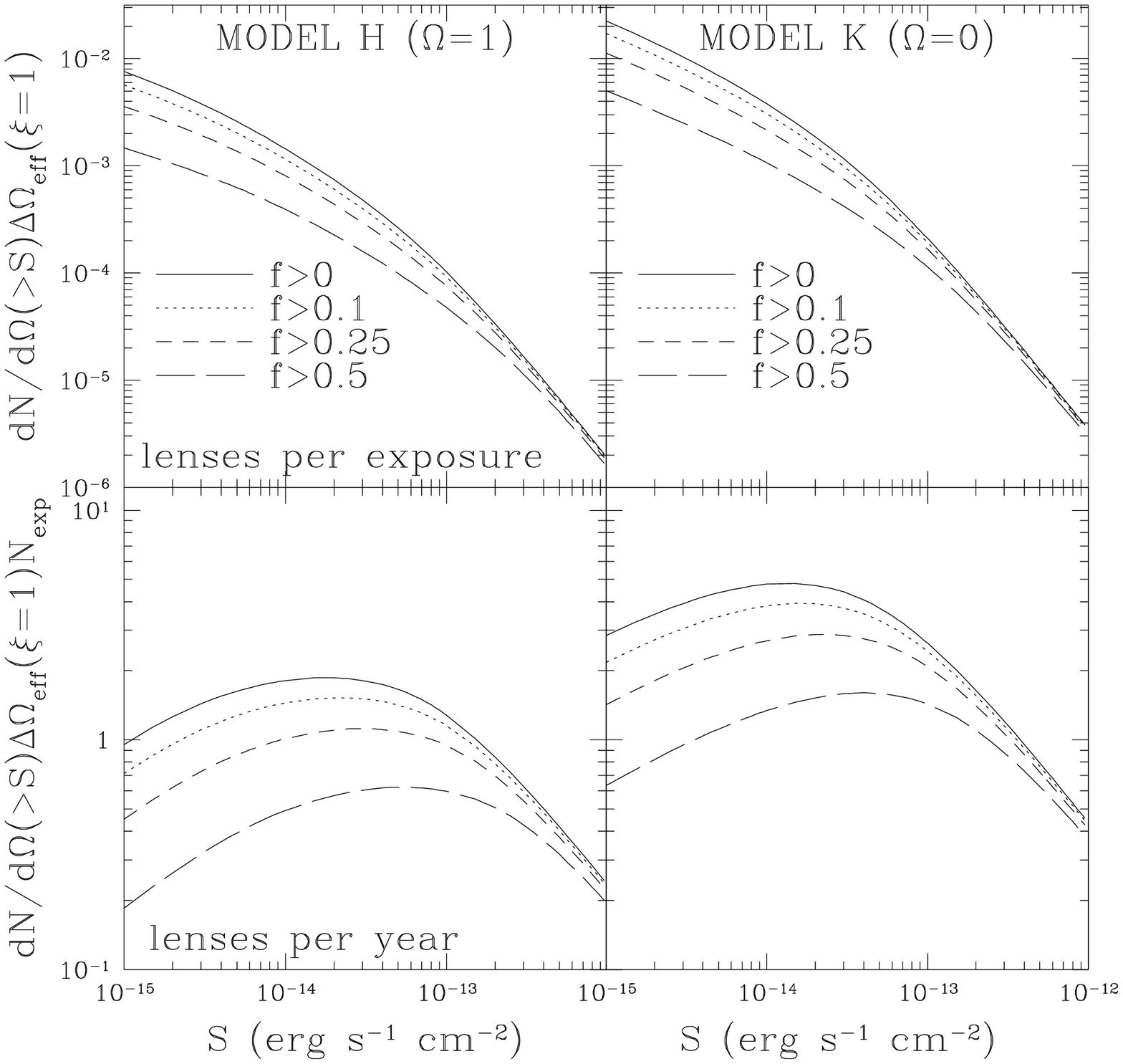,height=6.5in}}
\caption{(Top) The number of lensed X-ray sources per telescope
pointing including the effective area and flux ratio limits for
$\Omega_0=1$ and LF model H (left) and for $\Omega_0=0$ and LF model K
(right). (Bottom) The expected number of lenses per year of high
resolution imaging including selection effects. }
\end{figure}
 
Figure 1 shows the expected number of X-ray sources and
lensed X-ray sources per square degree as a function of flux assuming
a perfect detector ($f>0$ or $M_{min}=2$, and $C(x)=1$). Figure 2
shows the redshift distribution of the lensed and unlensed sources for
integrations of 1, 10 and 100~ksec assuming an exposure
time of $(S/2.5\times 10^{-13}\fluxb)^{-1}$~ksec for a 5--$\sigma$
point source detection (e.g., Jerius et al. 1997).  The lensing
probability peaks near $S=10^{-13}$~\fluxa, which provides the best
balance between magnification bias and source redshift.  The magnification
bias is highest for the brightest sources (steep number counts, far
from the break in number counts), while the lens cross section is highest
for faint sources (highest average redshift).  For brighter sources
the probability drops because of the low average source redshift and
for fainter sources it drops because of the flattening of the number
counts distribution.  The peak lensing probability of 0.2--0.4\%
(depending on the cosmological model) is lower than for bright quasars
(about 1\%) but higher than for radio sources (about 0.1--0.2\%).  The
total number of X-ray lenses is enormous, reaching roughly one per
square degree for $S>10^{-15}$~\fluxa.  Particularly for the
$\Omega_0=1$ model, the predictions are underestimates because the
luminosity function models underpredict the observed number counts of
sources (see Boyle et al. 1994).

Observational selection effects determine the fraction of these lenses
that can be found, so we next estimate the number of observable lenses
per HRC or ACIS exposure.  The fundamental problem with the Chandra
Observatory for conducting a lens survey is the strong variation in
the resolution with the distance from the field center.  We estimate (from
Kenter et al. 1997) that the radius encircling 50\% of the energy is
approximately $r_{50}=0\farcs5+6\farcs0(D/10\arcmin)^2$ at a distance
$D$ from the field center.  The minimum separation for recognizing
multiple images can be approximated by a small multiple of $r_{50}$,
$\Delta\theta_{min}=\xi r_{50}$ with $1 < \xi < 2$.  Thus, we can
define an effective area for the detection of multiply imaged X-ray
sources by
\begin{equation}
  \Delta\Omega_{eff} (\xi) = 2\pi \int_0^\infty  D dD C(\xi r_{50}/\Delta\theta_*)
\end{equation}
where $C(x)$ is the angular selection function introduced in eqn. (5).
We can use an upper limit to the integral of $\infty$ rather than the
physical detector size because the exponential cutoff in $C(x)$ makes
it unimportant.  For reasonable count rates the best estimate is
$\Delta\Omega_{eff}(\xi=1)=0.012$ square degrees, but if pessimistic,
$\Delta\Omega_{eff}(\xi=2)=0.0035$ square degrees.  Unfortunately, the
effective area of the detector is far smaller than its total area.
Figure 3 shows the expected number of lenses per telescope pointing
(i.e. in an area $\Delta\Omega_{eff}(\xi=1)$) for limits on the
detectable flux ratio of $f>0$, $0.1$, $0.25$ and $0.5$.  The effect
of the flux ratio limit is smallest for bright sources, where the 
magnification bias leads to a sample dominated by lenses with modest
flux ratios, and enormous for faint sources.  While the expected
number of lenses drops rapidly as we move to brighter sources, the
reduced exposure time needed to detect bright sources greatly
increases the number of possible exposures.  The number of exposures
that can be taken per year is roughly $N_{exp} = 10^4
(S/10^{-13}\fluxb)$, so the number of detectable lenses per year
of high resolution imaging ($\Delta\Omega_{eff}N_{exp}dN/d\Omega(>S)$)
is roughly independent of the flux limit (see Figure 3).  
If the Chandra Observatory
were devoted only to high resolution imaging, then we would expect to
find 1--3 lenses per year.

\section{CLUSTER LENSES}

\begin{figure}
\centerline{\psfig{figure=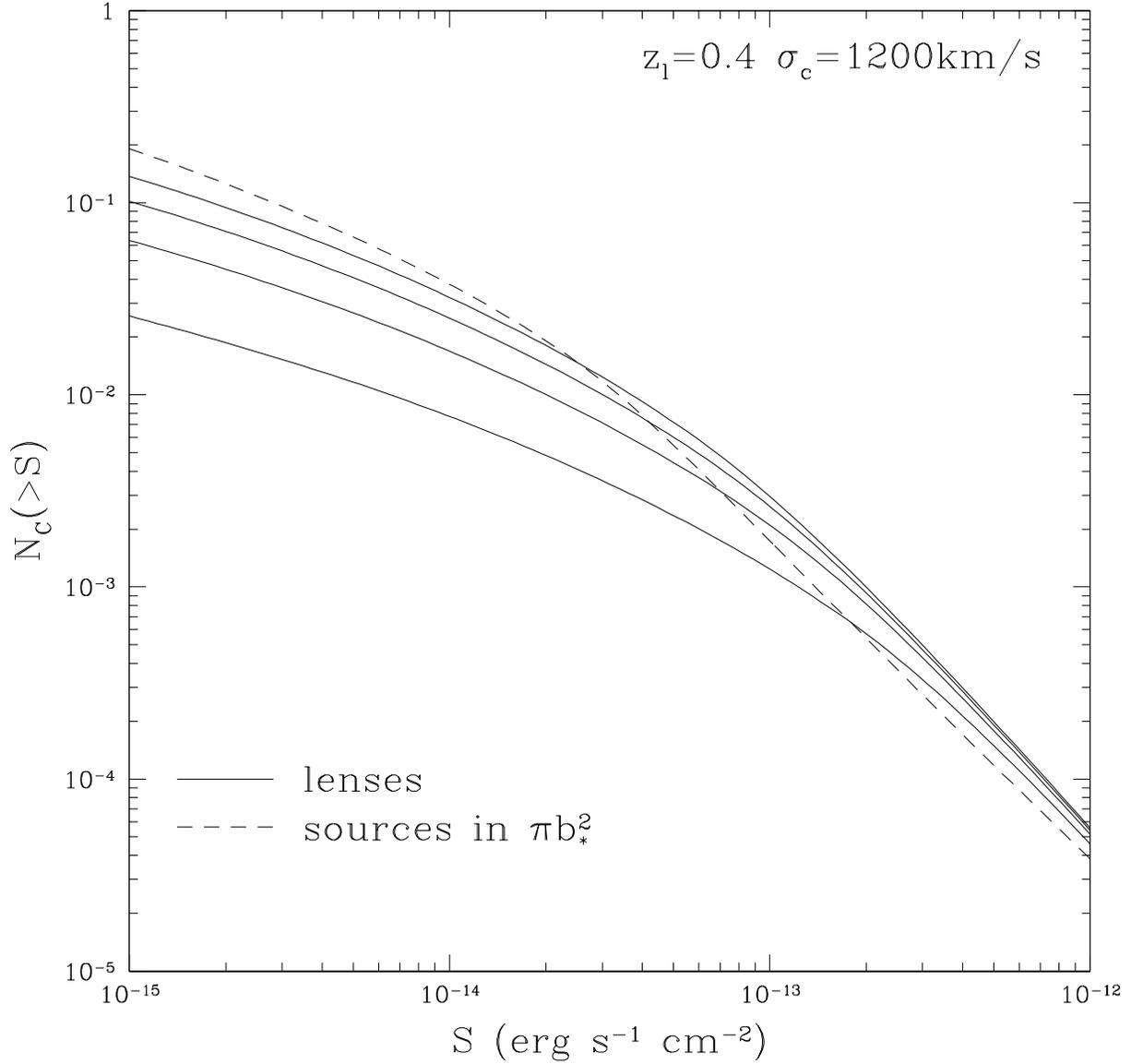,height=6.5in}}
\caption{The number of multiply-imaged X-ray sources as a function of
flux behind a typical cluster containing giant arcs, with $z_l=0.4$ and
$\sigma_c=1200$~km~s$^{-1}$.  The solid lines show the expected
number of lensed images brighter than flux $S$ with limits on the
detectable flux ratio of $f>0$, $0.1$, $0.25$ and $0.5$ (top to
bottom).  The dashed line shows the number of sources which would be
found within solid angle $\pi b_*^2$ in the absence of any lensing
effects.  }
\end{figure}
 
\begin{figure}
\centerline{\psfig{figure=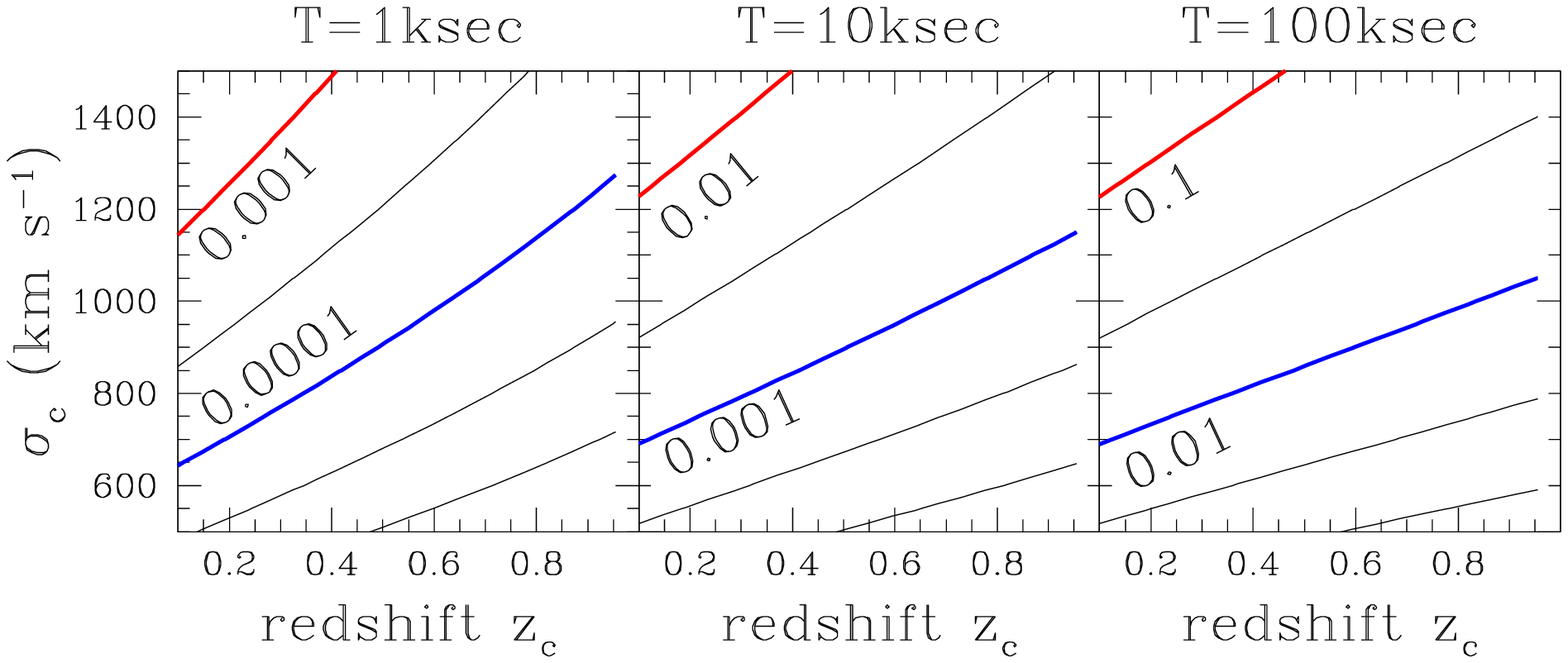,height=6.5in}}
\vspace*{-8cm}
\caption{Contours of the expected number of lenses with flux ratios
$f>0.5$ as a function of cluster redshift and velocity dispersion for
exposure times of 1~ksec (left), 10~ksec (middle), and 100~ksec
(right).  Contours are spaced at intervals of 0.5 dex (a factor of
$\sim 3$),
with labels on the heavy, shaded contours.  The increase in the number
of detections for weaker flux ratio restrictions can be determined
from Figure 4.}
\end{figure}

It is very unlikely to find a cluster acting as a lens in a randomly
selected field (see Kochanek 1995, Wambsganss et al. 1995, Flores \&
Primack 1996, Maoz et al. 1997) -- the high cross sections of clusters
compared to galaxies are far outweighed by their rarity.  However,
many Chandra observations will be centered on intermediate redshift
clusters, so they are pre-selected to have a massive, lensing object
in the field.  The critical radius scale for an SIS lens with velocity
dispersion $\sigma_c$ is $b_* = 4\pi (\sigma_c/c)^2$.  For a
particular lens and source redshift the image separation is
$\Delta\theta=2b_* D_{LS}/D_{OS}$ for distances from the lens
(observer) to the source of $D_{LS}$ ($D_{OS}$).  The multiple-imaging
cross section is $\tau_c(z_s) = \pi \Delta\theta^2/4$, so the expected
number of lenses behind a cluster of velocity dispersion $\sigma_c$
and redshift $z_l$ is
\begin{equation}
   N_c(>S) =  \int_{z_l}^\infty dV_s \tau_c(z_s) \int_{L_{min}}^\infty d L \int_{M_{min}}^\infty
        {d M \over M } { d P \over d M} { d N \over dL d z} \left( {L \over M } \right),
\end{equation}
where as before, $dV_s=D_{OS}^2 dD_{OS}$ for a flat cosmology.  The
expected number of lenses $N_c$ is very weakly dependent on the
cosmological model (because the cross section depends only on the
distance ratio $D_{LS}/D_{OS}$), so we restricted the calculation 
to $\Omega_0=1$ and luminosity function H.  Even so, the number of lenses is
underestimated because the LF model underestimates the number of faint
X-ray sources.  The image separations produced by a massive cluster
are sufficiently large to allow us to assume that no systems are lost
due to limitations in angular resolution, although we must still
impose limits on the detectable flux ratios.  Figure 4 shows the
number of lenses expected behind a typical ``giant-arc'' cluster 
(velocity dispersion $\sigma_c=1200$~km~s$^{-1}$) at redshift $z_c=0.4$
as a function of the image flux ratio limit $f$.  The
expected number of lensed sources is roughly equal to the number of
X-ray sources expected within solid angle $\pi b_*^2$ -- while the
average cross section is smaller than $\pi b_*^2$, the magnification
bias compensates.  Figure 5 shows the expected number of lenses found
in $1$, $10$ and $100$~ksec images of clusters as a function of their
redshift and velocity dispersion.  As in the serendipitous surveys,
individual observations are unlikely to detect multiply-imaged
systems, but the accumulated results of all imaging programs will
find lensed sources.  The number of lenses detected is of order 1--10 
for each year devoted to imaging clusters, depending on the mass and
redshift distributions of the clusters.  Whether the SIS is a realistic
representation of cluster lenses is an open question (e.g. see Williams,
Navarro \& Bartelmann 1999), but the cross section estimates should be 
approximately correct.

\section{SUMMARY}

The Chandra X-ray Observatory will discover both serendipitous lenses,
where a random background source is found to be lensed by a 
foreground galaxy, and cluster lenses, where a background
source is found to be lensed by a cluster that is the target of a
Chandra pointed observation.  The number of detectable systems is 1--3
serendipitous lenses and 1--10 cluster lenses per year of imaging
time, roughly independent of the flux limit of the observations and
including strong limits on the detectability of the lensed images.
These are probably {\it underestimates} because the Boyle et
al. (1994) luminosity functions we used for our calculations undercount
the numbers of X-ray sources at faint flux limits.  The X-ray Multi-Mirror
Mission (XMM, http://astro.estec.esa.nl/XMM), with its
coarser angular resolution (5\farcs0 FWHM), will be unable to detect
gravitational lenses produced by galaxies.  However, its high sensitivity will
make it very useful for detecting cluster lenses.

The total number of lensed X-ray sources is enormous, roughly
$(10^{-15}\fluxb/S)$ lenses per square degree brighter than a 
soft X-ray flux $S$, with none of the confusion
problems which interfere with searches for gravitational lenses in the
optical or radio.  An X-ray telescope with the resolution of the Chandra
Observatory over a wide field of view  would be an extraordinarily 
efficient instrument for finding gravitational lenses.  Alternatively,
deep, high resolution optical images of X-ray sources should be an
efficient means of searching for new gravitational lenses.

\acknowledgments 

Acknowledgments: We would like to thank Adam Dobrzycki for comments and for
producing simulated HRC images of gravitational lenses. We would also like to thank
Richard Mushotzky and Xavier Barcons for their comments.  CSK is supported by 
NASA ATP grant NAG5-4062.


\begin{references}

\reference{} 
Barkana, R., Leh\'ar, J., Falco, E. E., Grogin, N. A., Keeton, 
C. R. \& Shapiro, I. I. 1999, \apj, in press (astro-ph/9808096)

\reference{}
Bernstein, G. \& Fischer, P. 1999, \apj, in press (astro-ph/9903274)

\reference{} 
Boyle, B. J., Shanks, T., Georgantopoulos, I., Stewart, G. C. \& 
Griffiths, R. E., 1994, \mnras, 271, 639

\reference{}
Carroll, S. M., Press, W. H. \& Turner, E. L. 1992, \araa, 30, 499
 
\reference{}
Cheng, Y.-C. N. \& Krauss, L. M. 1999, astro-ph/9810393

\reference{}
Chiba, M. \& Yoshii, Y. 1999, \apj, 510, 42

\reference{}
Cooray, A.R., 1999, \aap, 342, 353
 
\reference{}
Fabbiano, G. 1989, \araa, 27, 87
 
\reference{}
Faber, S. \& Jackson, R. E. 1976, \apj, 204, 668

\reference{}
Falco, E.E., Kochanek, C.S. \& Mu\~noz, J.A., 1998, ApJ, 494, 47

\reference{}
Fassnacht, C. D. et al. 1999, \aj, 117, 658 

\reference{} Flores, R.A. \& Primack, J.R., 1996, ApJ, 457, 5L

\reference{}
Griffith, M., Heflin, M., Conner, S., Burke, B. \& Langston, G. 1991, 
\apjs, 75, 801

\reference{} Helbig, P. 1999, astro-ph/9904311

\reference{}
Helbig, P., Marlow, D., Quast, R., Wilkinson, P. N.,
Browne, I. W. A. \& Koopmans, L. V. E. 1999, \aaps, 136, 297

\reference{}
Impey, C. D., Falco, E. E., Kochanek, C. S., Leh\'ar, J., McLeod, B. A.,
Rix, H.-W., Peng, C. Y. \& Keeton, C. R. 1998, \apj, 509, 551

\reference{}
Jerius, D., Zhao, P., Van Speybroek, L., Tennant, A., Swartz, D., Schwartz,
D. A., Podgorski, W. A., Harris, B., Graessle, D. E., Gaetz, T. J., 
Freeman, M. D., Elsner, R., Edgar, R. J. \& Cohen, L. M. 1997, \baas, 190.2901

\reference{} 
Keeton, C. R., Kochanek, C. S., \& Falco, E. 1998, \apj, 509, 561

\reference{} Kenter, A.T., Chappell, J.H., Kobayashi, K., et al., 1997, Proc. SPIE, 3114, 26

\reference{} King, L. J. \& Browne, I. W. A. 1996, \mnras, 282, 67

\reference{} Kochanek, C. S., Falco, E. E., Impey, C. D., Leh\'ar, J. L., 
McLeod, B. A., \& Rix, H.-W. 1999, ``After the Dark Ages: When Galaxies Were Young'',
9th Maryland Astrophysics Conference, editors S. Holt and E. Smith,
AIP conference proceedings 470, page 163, (AIP: New York)

\reference{} Kochanek, C. S. 1996, \apj, 473, 595 

\reference{} Kochanek, C.S., 1995, \apj, 453, 545 

\reference{} Kochanek, C. S. 1993a, \apj, 417, 438  

\reference{} Kochanek, C. S. 1993b, MNRAS, 261, 453 

\reference{} Luppino, G. A., Gioia, I. M., Hammer, F., Le F\`evre, O. \&
Annis, J. A. 1999, \aaps, 136, 117

\reference{} Maoz, D., Rix, H.-W., Gal-Yam, A., \& Gould, A., 1997, ApJ, 486, 75

\reference{} Rix, H.-W., de Zeeuw, P.T., Cretton, N., van
        der Marel, R.P., \& Carollo, C.M. 1997, \apj, 488, 702

\reference{} Schechter, P. 1976, \apj, 203, 297

\reference{}
Schneider, P., Ehlers, J. \& Falco, E. E. 1992, in ``Gravitational Lenses'', 
Springer-Verlag, Berlin

\reference{} Turner, E.L., 1990, ApJ, 365, L43

\reference{} Wambsganss, J., Cen, R., Ostriker, J.P. \& Turner, E.L., 1995, Science, 268, 274

\reference{} Williams, L.R.L, Navarro, J.F. \& Bartelmann, M., 1999, astro-ph/9905134
\end{references}
\end{document}